\documentclass{aa}
\usepackage{float}

\usepackage{txfonts}
\usepackage{natbib}
\usepackage{color}
\usepackage{graphicx}
\usepackage{hyperref}
\usepackage{textcomp}

\usepackage{gensymb}           
\usepackage{siunitx}[=v2]      
\usepackage{multicol}
\usepackage{multirow}
\usepackage{longtable}

\bibpunct{(}{)}{;}{a}{}{,}

\newcommand{\Rsun}{\hbox{$\mathcal{R}_\odot$}}
\newcommand{\Msun}{\hbox{$\mathcal{M}_\odot$}}

\DeclareSIUnit\pc{pc}
\DeclareSIUnit\mag{mag}
\DeclareSIUnit\au{\astronomicalunit}

\newcommand{\dc}{\hbox{$\delta$~Cir }}
\newcommand{\de}{\hbox{$\delta$~Cir}}

\newcommand{\hp}{\hbox{H$_{\rm p}$}}

\begin{document}
   \title{$\delta$~Circini: A massive hierarchical triple system \\ with an eclipsing binary}
\titlerunning{\de}

\author{J.~\v{S}vr\v{c}kov\'a\inst{1,2}\and
        P.~Harmanec\inst{1}\and
        R.~Klement\inst{2}\and
        Th.~Rivinius\inst{2}\and
        B. N.~Barlow\inst{3}\and
        J.~Jury\v{s}ek\inst{4}\and
        M.~Ma\v{s}ek\inst{4}\and
        K.~Ho\v{n}kov\'a\inst{5}\and
        A.~Opli\v{s}tilov\'a\inst{1}
        \offprints{J. \v{S}vr\v{c}kov\'a,
        \email  svrckovaj@gmail.com}
}
\institute{
   Astronomical Institute of Charles University,
   Faculty of Mathematics and Physics, V~Hole\v{s}ovi\v{c}k\'ach~2, CZ-180 00 Praha~8,
   Czech Republic, \email  {svrckovaj@gmail.com}
\and
   European Organisation for Astronomical Research in the Southern Hemisphere (ESO), Casilla 19001, Santiago 19, Chile
\and
 Department of Physics and Astronomy, University of North Carolina at Chapel Hill, Chapel Hill, NC 27599, USA
\and
 FZU - Institute of Physics of the Czech Academy of Sciences, Na Slovance 1999/2, Prague, 182 21, Czech Republic
 \and
 Variable Star and Exoplanet Section of the Czech Astronomical Society, Fri\v{c}ova 298, 
 251 65 Ond\v{r}ejov, Czech Republic
}

   \date{Release \today}

\abstract{$\delta$ Circini is known to be a massive multiple system containing a 3.9~d inner eclipsing binary in a~slightly elliptical orbit exhibiting slow apsidal motion and a distant tertiary with a probable period of 1644~d. All three components of the system are O- or B-type stars. 
We carried out a comprehensive study of the system, based on light curves from TESS and other instruments, a new series of echelle spectra, older spectra from the ESO archive, and several VLTI interferometric observations. Due to the large amount of different types of data covering both orbits in the system, we obtained a more precise value of the long orbital period ($1603.24\pm0.19$ d) and fully determined all other orbital parameters.
Although both orbits are eccentric, their period ratio is large enough for the system to be dynamically stable. The inner and outer orbits are in the same plane, which means that no Kozai-Lidov mechanism is acting in the system. 
Assuming solar metallicity in our MESA models, we found ages of $(4.4\pm 0.1)$, $(4.7\pm 0.2)$, and $(3.8\pm1.3)$\,Myr for the primary, the secondary, and the tertiary, respectively. Our evolutionary scenario predicts that the inner eclipsing binary will merge within approximately 1.7\,Myr and eventually evolve into a black hole.
The distance to the system, estimated from the angular size of the outer orbit is $(809.9 \pm 1.8)$ pc, which implies that $\delta$~Cir might be located close to the centre of a stellar population ASCC 79, a subgroup of the young Circinus complex. With a total mass of ($53.04\pm0.29$) \Msun, $\delta$~Cir can contribute a~significant fraction of the total mass of the population.
}


   \maketitle

\section{Introduction}
Even though massive OB stars spend only several million years on the main sequence and make up only a small fraction of the stellar population in our galaxy, they have a very important influence on the Universe. With large luminosities, strong UV radiation, stellar winds, and heavy chemical elements released in supernova explosions, they heat the interstellar medium in the surroundings, ionise it, and change its chemical composition. 
Massive stars are often found in multiple star systems: around 90\% of them have at least one companion \citep{Pauwels2023}. Stars with companions may evolve in a very different way than single stars. Due to their evolutionary expansion, massive binaries with periods up to 1500~d can undergo large-scale mass exchange, and for about 40\% of O stars this already occurs during their evolution on the main sequence \citep{Sana2012}. Only a small fraction of massive stars evolve towards a supernova explosion without any disruption. The properties of multiple star systems depend on their formation and the interactions between their components. Therefore, these systems can be used to test theories of the formation and the evolution of massive stars. In addition to $\delta$~Cir, discussed in this study, we are aware of several other relatively compact multiple OB systems. One is IU Aur \citep{iuaur2003} composed of an OB binary with a period of 1\fd8115 orbiting in a 293\fd3 orbit with a third body, which could possibly also be a close binary. Other
notable systems are HD~152246 \citep{nasseri2014}, a 6\fd0049 binary in a highly eccentric 470\fd0 orbit with a third body; $\xi$~Tau \citep{Nemravova2016}, a hierarchical quadruple system composed of a 7\fd14 eclipsing binary in a~145~d orbit with a rapidly rotating B tertiary and in a 51~yr orbit with an F body; and, possibly, HD~92206C \citep{mayer2017}, a~2\fd0225 binary with a possible tertiary suspected from the $O-C$ residuals. 

$\delta$~Circini~(HR 5664, HD~135240, CPD~$-60^\circ5701$, HIP~74778) is a bright OB multiple system containing a 3.9~d eclipsing binary. The determination of its accurate masses and radii has long been complicated by the fact that its complete light curve was not easy to obtain from a single site. A detailed study of the object, where its observational history is also summarised, was published by \citet{mayer2014}. These authors have demonstrated that the observed spectrum is a combination of three spectra, two of them belonging to the eclipsing 3.9~d binary, and the third to a distant tertiary. Analysing the variation of the systemic velocity of the eclipsing binary with time, as well as direct radial-velocity (RV) measurements of the third star, they demonstrated that the binary and the tertiary revolve around a common centre of gravity in a highly eccentric orbit with a 1644~d period. This, however, needed further verification. By modelling a historical light curve and a light curve from Hipparcos {\hp} observations \citep{esa97}, they derived probable basic physical elements of the system. They also speculated that the tertiary is likewise a binary system, decomposing its lines in a few spectra into two components and further supporting this by the large residuals of the measured tertiary RVs from the orbital solution.

Since then, good light curves with complete phase coverage have been obtained with the TESS space telescope and a ground-based telescope named FRAM. \citet{southworth2022} analysed the TESS photometry in a search for the presence of pulsations but did not find any significant frequency. Based on the work by \citet{shi2022}, the TESS light curve of $\delta$~Cir shows an eccentric eclipse and similar pulsating variations; however, they could not find any pulsation frequency. According to \citet{Kerr2025}, the system could possibly be a member of the young population ASCC 79 in the Circinus complex, a region with dense molecular gas and active star formation.

To elaborate on the findings of \citet{mayer2014}, we obtained a series of high-resolution echelle spectra with the CHIRON spectrograph and downloaded public HARPS and FEROS spectra from the ESO archive. We also downloaded publicly available near-IR interferometric data from the PIONIER and GRAVITY instruments. A detailed analysis of these new observations, as well as existing RV sets, already used by \citet{mayer2014}, is the subject of this study.

\section{Observational data}

\subsection{Spectroscopy}

We downloaded 17 FEROS and 95 HARPS reduced spectra from the ESO Science Archive. FEROS \citep{FEROS1999} is a fibre-fed high-resolution ($R \sim 48 000$) echelle spectrograph attached to the MPG/ESO 2.2m telescope located at the ESO La Silla Observatory. The FEROS spectra are from 2007-2009 (HJD 2454277.5--2454302.5). HARPS \citep{HARPS2003} is a high-resolution fibre-fed echelle spectrograph attached to the ESO 3.6m telescope at the La Silla Observatory in Chile. The resolution of its spectra is up to $R \sim 115 000$. The HARPS spectra of \dc were observed during the years 2009-2012 (HJD 2455003.4--2456136.5). 

In addition, we used 39 echelle spectra from the CHIRON spectrograph mounted on the SMARTS 1.5m telescope at the Cerro Tololo Interamerican Observatory (CTIO) \citep{CHIRON2013}. Of these, 11 were observed during January and February 2021 (HJD 2459219.9--2459294.8), and the rest were observed from February until April 2022 (HJD 2459615.9--2459692.7). The spectra were taken in fibre mode with spectral resolution of $R \sim 25000$. Their initial reduction was carried out through a standard pipeline from the observatory. Further steps, such as normalisation of individual echelle orders, cleaning of cosmic rays and flaws were done in the programme reSPEFO written by A.~Harmanec\footnote{\url{https://astro.troja.mff.cuni.cz/projects/respefo}\,.}. 

An interference pattern is present in 13 of the HARPS spectra. The fringes are quasiperiodic, and their intensity is comparable with some of the weaker helium lines. \citet{mayer2014} managed to remove the fringes from a short segment of the spectra; however, we were unable to find the corrected spectra in our data collection. In the end, we decided to use only the 82 spectra without interference. Furthermore, four of the CHIRON spectra were over-exposed and, for this reason, were completely unusable for our purpose. We also discarded one FEROS and two HARPS spectra with a signal-to-noise ratio lower than 100.

\subsection{Photometry}

Two of the light curves we used were observed by the Transiting Exoplanet Survey Satellite (TESS). This satellite is equipped with an instrument consisting of four optical cameras with a wide field-of-view ($24\degree$ × $24\degree$), 105 mm aperture and focal ratio $f/1.4$ \citep{Ricker2015}. TESS observes the sky in sectors, each of them 27.4 days long. The $\delta$ Cir data come from sectors 12 and 65. The light curves are released by the TESS
Science Processing Operations Centre (SPOC) and archived at the Mikulski Archive for Space Telescopes (MAST\footnote{\url{https://mast.stsci.edu/portal/Mashup/Clients/Mast/Portal.html}}). We downloaded the standard data product, which is the simple aperture photometry (SAP). We binned the data to get a lower number of data points (roughly 1350 in each sector) to reduce the time necessary for modelling the light curves.

\begin{table}[h!]
\centering
\caption{\label{data_phot}Journal of the photometric observations of \de.}
\begin{tabular}{rccr}
\hline\hline\noalign{\smallskip}
Instrument & HJD$-$2400000 &Passband&No. of   \\
           &             &        &nights   \\
\noalign{\smallskip}\hline\noalign{\smallskip}
Hipparcos & 47918.4--49042.6 & \hp      & 24\\
FRAM      & 57019.8--57145.7 & $V$      &126\\
TESS S12  & 58625.0--58652.9 & TESS     & 27.4\\
TESS S65  & 60068.8--60096.6 & TESS     & 27.4\\
\noalign{\smallskip}\hline\noalign{\smallskip}
\end{tabular}
\end{table}

\begin{table}[h!]
\centering
\caption{\label{data_spec}Spectroscopic observations of \de.}
\begin{tabular}{rcc}
\hline\hline\noalign{\smallskip}
Instrument & HJD$-$2400000 & No. of spectra  \\
\noalign{\smallskip}\hline\noalign{\smallskip}
FEROS & 54277.5--54302.5 & 17\\ 
HARPS & 55003.4--56136.5 & 95\\
CHIRON & 59219.9--59692.7 & 39\\
\noalign{\smallskip}\hline\noalign{\smallskip}
\end{tabular}
\end{table}

Another light curve we used was obtained by a robotic Schmidt-Cassegrain telescope FRAM, located at the Pierre Auger Observatory in Argentina. This light curve was obtained with the Johnson $V$ filter and a $f/2.8$ photographic lens attached to the main tube of the telescope. The focal length of the lens is \SI{300}{mm}. Together with a G4-16000 CCD camera, it is used for wide-field imaging. 
The last light curve that we used is the wide-passband H$_{\rm p}$ photometry obtained by the Hipparcos satellite, equipped with an eccentric Schmidt telescope with an aperture of \SI{0.29}{m}. 
All the photometric observations and spectra used in this work are summarised in Tables~\ref{data_phot} and \ref{data_spec}.

\subsection{Interferometry}

In total, 12 interferometric observations of \dc from the Very Large Telescope Interferometer (VLTI) were used in our analysis. Instruments on the VLTI combine beams from four telescopes, either the movable 1.8m Auxiliary Telescopes (AT) or the 8.2m Unit Telescopes (UT). Of the 12 observations, all were taken with the AT array in different versions of the `Large' configuration with a maximum baseline of $\sim120$\,m (see Notes in Table~\ref{tab:interferometry} for the exact telescope configuration for each observation). Nine of the observations come from the beam combiner PIONIER \citep{PIONIER2011}, which works in the H band (1.51 to \SI{1.76}{\um}) with low spectral resolution ($R \sim 40$). We downloaded the PIONIER data from the Optical Interferometry Database\footnote{\url{http://oidb.jmmc.fr}} \citep[OIDB,][]{Haubois2014}, already reduced and calibrated. The extracted observables were six sets of visibility-squared data points (V2), one per pair of telescopes, and four sets of closure phases (T3PHI), one per each telescope triangle.

The rest of the interferometric observations are from the GRAVITY beam combiner \citep{GRAVITY2017}, which operates in the K band (1.98 to \SI{2.40}{\um}) with spectral resolution reaching up to $R\sim4000$. For the public \dc observations, the instrument was used in single-field on-axis mode and with high spectral resolution of $R\sim4000$. Each of the GRAVITY observations was done in a sequence of object-sky-object, which resulted in two sets of data that are slightly different from each other. This is caused by the change in baseline projection due to the rotation of the Earth. All data were obtained in combined polarimetry mode. GRAVITY observations were reduced and calibrated with the official ESO reduction pipeline, version 1.6.6, in the ESO Reflex environment \citep{ESOReflex2013}. The extracted interferometric observables for each dataset were six sets of absolute visibilities (|V|), six sets of differential phases (DPHI), and four sets of T3PHI. In addition, the GRAVITY data also include a full $K$-band spectrum at the same resolution of $R\sim4000$.

\section{Analysis of the interferometric data}

For the visualisation and analysis of the interferometric data, we used the Python code PMOIRED \citep{Merand2022}. PMOIRED enables fitting interferometric data with geometrical models consisting of multiple components such as uniform disks, Gaussians, and Keplerian disks. The model parameters were obtained via $\chi^2$ minimisation. Additional PMOIRED capabilities include the fitting of spectro-interferometric features (spectral lines), telluric correction of GRAVITY high-resolution spectra, a specialised grid search algorithm to detect companions in multiple star systems, and a bootstrapping resampling algorithm that is used to determine realistic uncertainties of the fitted parameters. 

Our interferometric observations have angular resolution $\lambda / 2 B_{\rm max}$ (corresponding to the longest baseline $B_{\rm max}\sim120$\,m) of $\sim1.9$\,mas in the $K$ band (GRAVITY) and $\sim1.4$\,mas in the $H$ band (PIONIER). At this angular resolution, both the diameters of the individual stars ($\lesssim0.1$\,mas), as well as the separation of the inner binary ($\sim0.2$\,mas) are unresolved. On the other hand, in all of our datasets, we detect a clear binary signal caused by the presence of the third component. In the high-spectral-resolution GRAVITY data, we could not identify any prominent $K$-band spectral features, and we therefore re-binned the data to a lower spectral resolution of $\sim 100$ to speed up the grid search for the outer companion.

We used PMOIRED to search for the tertiary companion, fitting the T3PHI observable together with either V2 (PIONIER) or |V| (GRAVITY). Since the inner binary is unresolved, we used a simple model of two unresolved points, one representing the inner eclipsing binary and the other representing the third outer component of the system. The free parameters of the fit were the relative angular separation between the two points, the direction given by the position angle (measured from north to east), and the flux ratio of the outer component and the inner binary (in the $H$ band for PIONIER or $K$ band for GRAVITY). The grid search successfully converged to reasonable parameter values for each observing epoch with the reduced $\chi^2$ ranging from 0.3 to 1.9. Afterwards, we used the bootstrapping algorithm to obtain the final uncertainties of the fitted parameters. The list of all interferometric observations together with the relative astrometric positions fitted is in Table~\ref{tab:interferometry}, and the interferometric calibrators used for the observations are given in Table~\ref{tab:interferometry2}.

\begin{table*}
\caption{\label{tab:interferometry}Flux ratios and relative astrometric positions determined for each epoch of interferometric observations.}
\centering
\begin{tabular}{c c c c c c c c c c c}
\hline\hline \noalign{\smallskip}
Date & RJD & $f$ & $\rho$ (mas) & PA (\degree) & $\sigma_a$ (mas) & $\sigma_b$ (mas) & $\sigma_\mathrm{PA}$ (mas) & Instrument & Calibrator \\   \hline \noalign{\smallskip} 
2012 Jun 11 & 56090.0535 & $0.2016 \pm 0.0014$ & 3.790 & 131.82 & 0.0489 &	0.0129 & -30.8 & PIONIER & -\\
2012 Jun 14 & 56093.1194 & $0.1839 \pm 0.0042$ & 3.709 & 133.28 & 0.0112 &	0.0082 & -36.7 & PIONIER & -\\
2012 Jun 14 & 56093.1347 & $0.1702 \pm 0.0064$ & 3.728 & 134.33 & 0.0279 &	0.0132 & -0.8 & PIONIER & -\\
2014 Jul 29 & 56867.9809 & $0.2070 \pm 0.0033$ & 14.406 & -108.64 & 0.0291 & 0.0192 & -33.3 & PIONIER & 1\\
2016 Jul 28 & 57598.0239 & $0.2368 \pm 0.0024$ & 5.546 & 103.65 & 0.0088 & 0.0079 & -5.5 & PIONIER & 1\\
2017 Mar 12 & 57825.3668 & $0.1997 \pm 0.0029$ & 4.063 & -164.06 & 0.0188 & 0.0081 & -80.6 & GRAVITY & 2\\
2017 Apr 27 & 57871.3095 & $0.2006 \pm 0.0015$ & 4.886 & -149.95 & 0.0084 &	0.0051 & 40.8 & GRAVITY & 2\\
2017 Apr 28 & 57872.1865 & $0.1969 \pm 0.0013$ & 4.964 & -149.77 & 0.0163 & 0.0061 & 12.2 & PIONIER & 1\\
2017 Apr 30 & 57874.2998 & $0.1856 \pm 0.0045$ & 4.937 & -148.35 & 0.0280 & 0.0263 & 66.2 & GRAVITY & 2\\
2017 Jun 21 & 57926.0002 & $0.2420 \pm 0.0490$ & 6.129 & -138.41 & 0.0927 & 0.0425 & -18.3  & PIONIER & 1\\
2018 Apr 18 & 58227.1445 & $0.1946 \pm 0.0015$ & 12.505 & -116.54 & 0.0213 & 0.0095 & -41.8 & PIONIER & 1\\
2022 Apr 05 & 59675.1312 & $0.1899 \pm 0.0077$ & 9.563 & -124.21 & 0.0470 & 0.0115 & -48.9 & PIONIER & 3,4\\
\hline
\end{tabular}
\tablefoot{$f$ is the flux ratio of the third star and the inner binary, $\rho$ is the angular separation of the tertiary from the binary, PA is the position angle (measured from north towards east), $\sigma_a$ and $\sigma_b$ are the major and minor axis of the error ellipse and $\sigma_\mathrm{PA}$ is the position angle of the error ellipse major axis. The three observations in 2012 were done in telescope configuration K0-A1-G1-I1, the observation in 2014 was done in K0-A1-G1-J3 and the rest used configuration A0-G1-J2-J3 (all corresponding to `Large' configuration of the VLTI AT array). We were unable to find the calibrators of the first three observations. Detailed information about the rest of the calibrators is in Table \ref{tab:interferometry2}.}
\end{table*}

\begin{table}
\caption{\label{tab:interferometry2}Additional information about the calibrators of the interferometric observations.}
\centering
\begin{tabular}{c c c c c c c}
\hline\hline \noalign{\smallskip}
N. & Calibrator & Angular Diameter\tablefootmark{a} & Spectral Type \\   \hline \noalign{\smallskip} 
1 & HD\,136057 & 0.552 & K3\,III \\
2 & HD\,134449 & 0.484 & K0\,III \\
3 & HD\,135017 & 0.766 & K3\,III \\
4 & HD\,135693 & 0.435 & K3\,(III) \\
\hline
\end{tabular}
\tablefoot{
\tablefoottext{a}{Uniform disk diameter in the $H$ band for PIONIER data and in the $K$ band for GRAVITY data.}
}
\end{table}

\section{Disentangling of the spectra}
The lines of all components are wide and blend into each other. Thus, determining the RVs without assuming anything about the orbit of the stars is very difficult, and it results in less accurate RV values, especially near the conjunctions of the close binary, when the strength of the lines also varies due to eclipses. Thus, we decided to disentangle all the spectra, using the KOREL programme \citep{Hadrava1995,Hadrava2004}. KOREL uses the Fourier transform to disentangle the observed spectrum of the whole system into spectra of individual components, measure RVs, and determine orbital parameters of the system - all together as the best fit of the observed spectra. It can also work with the line profile variations that occur during eclipses.

For the disentangling, we assumed that $\delta$~Cir is a hierarchical triple system. First, we fitted the astrometric orbit (see Sect. \ref{outer-orbit}) and used the following fitted parameters as an initial estimate when disentangling the spectra: the orbital period (\SI{1603.7}{d}), eccentricity (\num{0.524}), argument of periastron (\SI{300.5}{\degree}), and time of periastron passage (RJD \num{57298}). The rate of the apsidal precession was estimated from preliminary models of the eclipsing binary (see Sect. \ref{inner-orbit}). The rest of the initial orbital parameters for KOREL was based on the paper by \citet{mayer2014}. We chose several shorter segments of the spectra containing helium lines and the H$\beta$ line and disentangled them separately. Specifically, the disentangled helium lines were \ion{He}{ii}~4686~\AA, \ion{He}{i}~4713~\AA, \ion{He}{i}~4922~\AA, \ion{He}{i}~5016~\AA, \ion{He}{ii}~5411~\AA, and \ion{He}{i}~6678~\AA. We kept the period of the outer orbit fixed; otherwise, the solution did not converge properly and only arrived at a local minimum. The triple-star model of the system in KOREL fits the data well. The residuals of the RVs are in most cases less than \SI{1}{\km\per\s}, which is a fairly small number when taking into account the wide blended lines in its spectrum.

A few of the spectra were taken during the eclipses, though their number is small compared to the total amount of spectra. Since KOREL can simulate the changes in the line profile that occur during eclipses, we decided to keep them for the disentangling. To verify that KOREL performed well even without discarding the eclipse spectra, we did a trial disentangling where we did not include these spectra and checked that we obtained basically the same orbital parameters and RVs. Moreover, the RV residuals corresponding to the eclipse spectra were not significantly larger than the other RV residuals. Since we independently disentangled seven short spectral regions, we had seven sets of RVs. We used their mean values in our further analysis and estimated the uncertainties as their rms. In comparison, the RV values of the third component obtained by fitting with Gaussian in the work of \citet{mayer2014} differ by up to several tens of \SI{}{\km\per\s} from our KOREL RVs.

\begin{figure}
  \resizebox{\hsize}{!}{\includegraphics{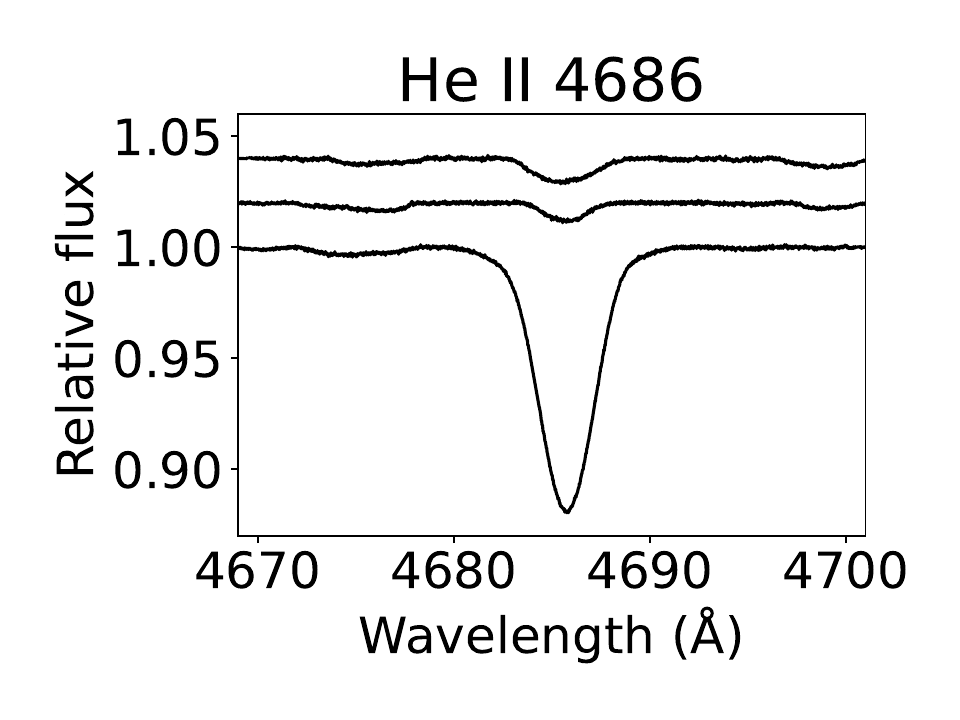}\includegraphics{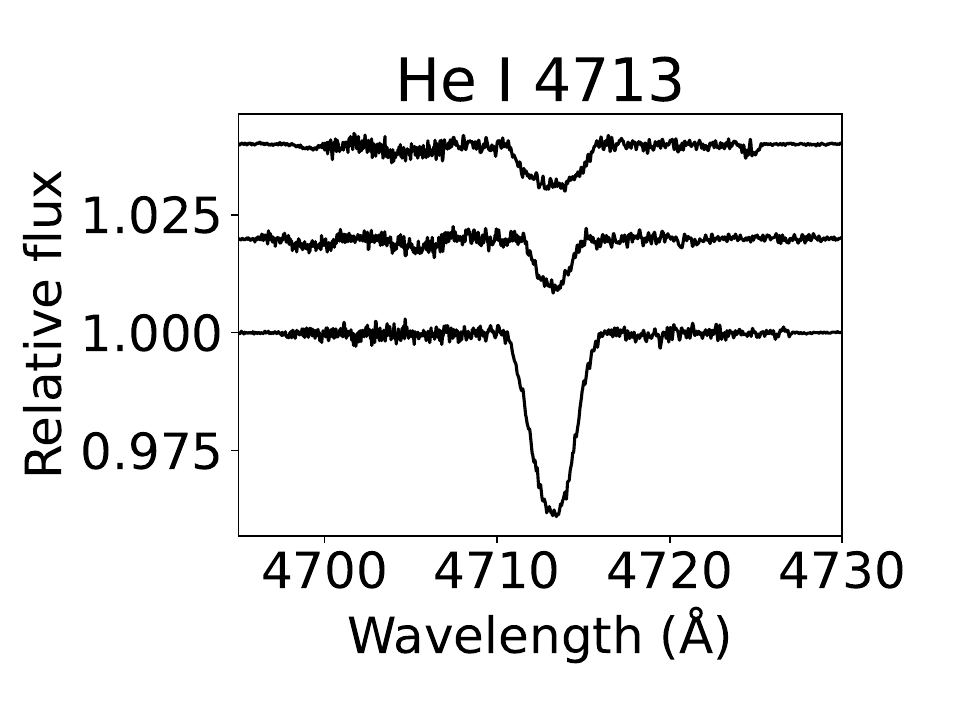}}
  \resizebox{\hsize}{!}{\includegraphics{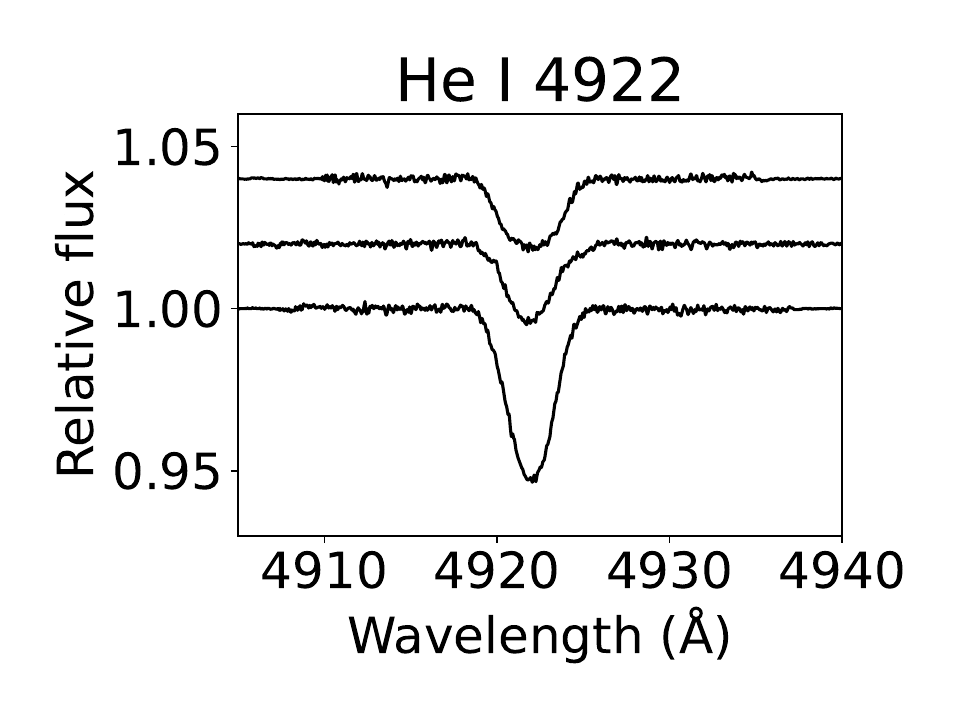}\includegraphics{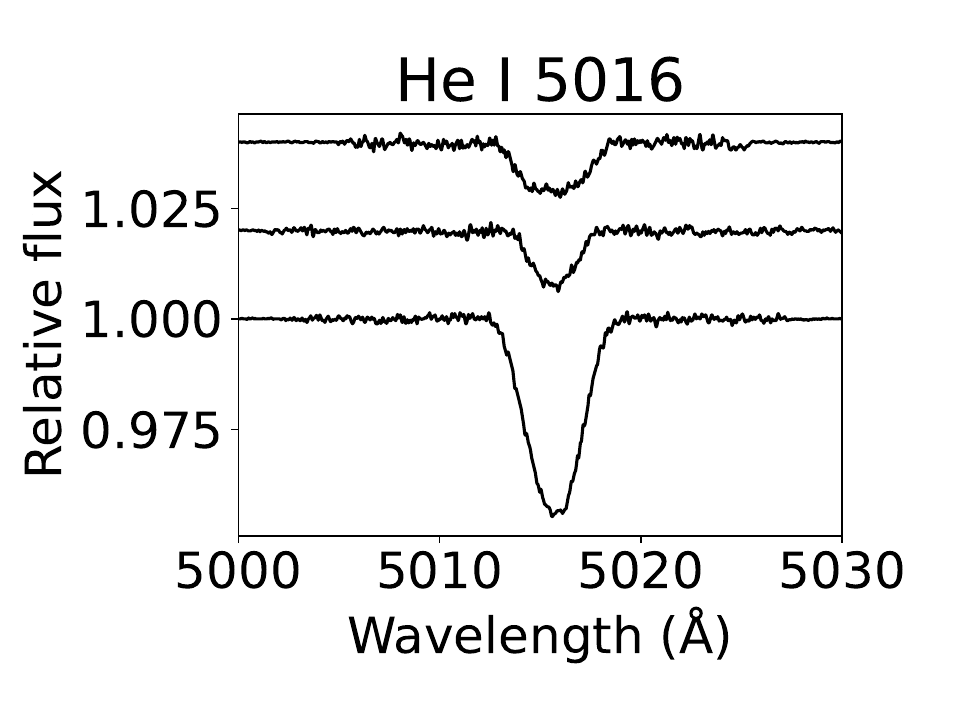}}
  \caption{Some of the normalised disentangled line profiles. Primary is always on the bottom, secondary in the middle and tertiary on the top. The continuum level of the secondary and the tertiary is shifted up by 0.2 and 0.4 respectively.}
  \label{rozklad_spektier}
\end{figure}

We divided the disentangled RV values into two sets that describe the movement in the inner and outer orbit. We used the mass ratio of the 3.9~d binary to find the RV of its centre of mass; in other words, the RVs due to its motion on the long, 1603.7~d orbit. Then we subtracted the centre-of-mass motion from the disentangled RVs to get only values that describe motion on the short orbit. The error caused by the uncertainty of the mass ratio was, in most cases, considerably smaller than the uncertainty of the RVs themselves. The systemic velocity $v_\gamma$ is subtracted from the RVs in the output of KOREL. We got $v_\gamma=\SI{3}{\km\per\s}$, which is in agreement with \citet{mayer2014}. We added this value back to the RVs of the individual components.

Since KOREL applies Fourier transformations to the spectra during the disentangling, it can create wavy patterns in the continuum of the decomposed spectra. The patterns were the most prominent in the H$\beta$ line, which is located at the edge of two echelle orders of CHIRON spectra. Compared to the primary and the secondary, the continuum waves were larger in the disentangled lines of the tertiary. Thus, it was necessary to manually re-normalise the final decomposed spectra. In some cases, it was quite difficult to see where the undulating continuum ends and where the lines wings start. Some of the line profiles are shown in Fig.~\ref{rozklad_spektier}. 

We also fitted the normalised decomposed lines in the programme PYTERPOL \citep{Nemravova2016}, to obtain the effective temperatures, rotational velocities, $\log g$, and relative luminosities of the three components. The lines were compared to synthetic spectra from the OSTAR and BSTAR grids \citep{Lanz2003, Lanz2007}. In our case, the parameter we were most interested in was the temperature of the primary, which we needed to model the light curve of the inner eclipsing binary. The summary of the final parameters is given in Table \ref{tab:pyterpol}. We estimated the uncertainties based on the results of several independent fitting trials.

\begin{table}
\centering  
\caption{Stellar parameters obtained from fitting the disentangled lines in PYTERPOL.} 
\label{tab:pyterpol}              
\begin{tabular}{l l l l}    
\hline\hline \noalign{\smallskip}           
Parameter & Primary & Secondary & Tertiary\\   
\hline \noalign{\smallskip}
$T_\mathrm{eff}$ (K) & $33750\pm300$ & $27500\pm1000$ & $30250\pm500$ \\
$\log g$ & $3.8\pm0.1$ & $4.05\pm0.15$ & $3.4\pm0.1$ \\
$v_\mathrm{r} \sin i$ (\SI{}{\km\per\s}) & $117\pm5$ & $96\pm10$ & $152\pm10$ \\
\hline                    
\end{tabular}
\end{table}

The previous studies of $\delta$ Cir adopted effective temperatures corresponding to the then known spectral classifications. \citet{Penny2001} used values $T_1 = \SI{37 500 \pm 1 500}{\K}$, $T_2 = \SI{30 000 \pm 1 000}{\K}$, and $T_3 = \SI{29 000 \pm 2 000}{\K}$. Our results are closer to the temperatures used in the work of \citet{mayer2014}; $T_1 = \SI{34 000}{\K}$, $T_2 = \SI{29 000}{\K}$, and $T_3 = \SI{28 000}{\K}$. Both of these studies assumed that the tertiary star is the coldest, while our line fitting in PYTERPOL shows that the coldest component of the system is actually the secondary.

\section{Orbital analysis}
\subsection{Modelling the outer orbit}\label{outer-orbit}

\begin{figure}
  \resizebox{\hsize}{!}{\includegraphics{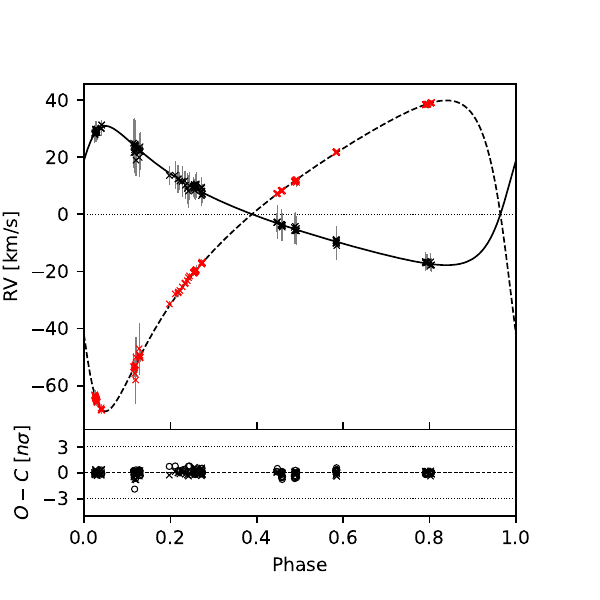}}
  \caption{\textit{Upper:} RV curve of the outer orbit. Black x-signs with grey error bars correspond to the RVs of the inner eclipsing binary centre of mass, while the red x-signs with grey error bars (in some cases too small to be seen) are the RVs of the outer component. The systemic velocity, $v_\gamma$, has been subtracted from the RVs. \textit{Lower:} $O-C$ residuals in the units of $\sigma$ for each individual point.
  }
  \label{RV_long_orbit}
\end{figure}

\begin{figure}
  \resizebox{\hsize}{!}{\includegraphics{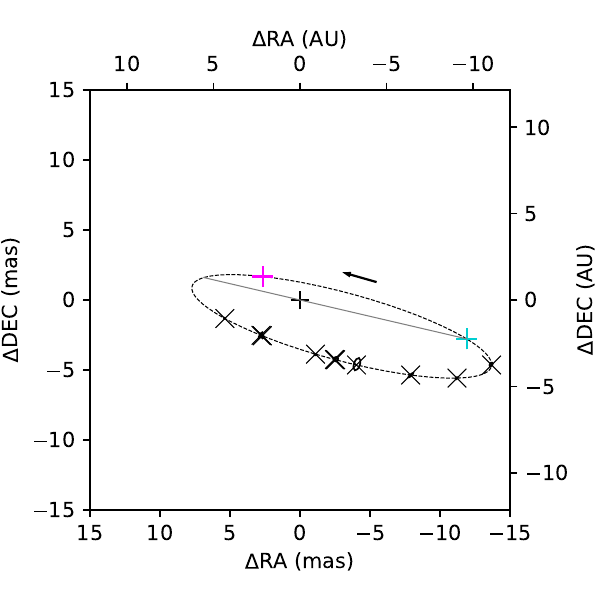}}
  \caption{The relative astrometric orbit of the outer component (dashed line) relative to the centre of mass of the inner binary (black plus sign). The black error ellipses (5$\sigma$) are our measurements (in some cases too small to be seen), and the black x-signs are the corresponding points on the computed orbit, which in each case fall within the error ellipses. The grey line is the line of nodes, the turquoise plus sign is the ascending node, the magenta plus sign is the periastron, and the arrow shows the direction of the outer orbit.}
  \label{astrometry_long_orbit}
\end{figure}

To obtain the parameters of the outer orbit, we used the IDL programme orbfit-lib\footnote{\url{https://www.chara.gsu.edu/analysis-software/orbfit-lib}}, which provides a simultaneous orbital fit of the astrometric and spectroscopic data and uses the Newton-Raphson method \citep{Schaefer2016}. The orbital parameters are summarised in Table \ref{tab:long_orbit}. Figs.~\ref{RV_long_orbit} and \ref{astrometry_long_orbit} show the RV curves and the astrometric orbit.

For comparison, we also used the Python spinOS\footnote{\url{https://github.com/matthiasfabry/spinOS}} programme \citep{Fabry2021, Fabry2021b}, where we chose the implemented Markov chain Monte Carlo sampling (MCMC) method. In case of an SB2 with available RVs for both stars in addition to relative astrometric positions, the parameter space in spinOS is overdetermined, as there are two possible ways to obtain the total mass of the system. To solve this issue, we fixed an arbitrary value of distance and fitted all the other parameters. All the fitted values with the exception of the total mass were close to expectations; therefore in the next round of fitting, we used the value of total mass based on the RV curve instead of the third Kepler law (and hence incorrect distance). Next, we fitted only for the distance of the system. We kept the value of total mass fixed also during the MCMC. All the posterior distributions were well defined. As we can see in Table \ref{tab:long_orbit}, the final set of spinOS parameters is in agreement within uncertainties with the parameters obtained from orbfit-lib. The parameter uncertainties given by MCMC in spinOS are considerably larger than those estimated by orbfit-lib. Since the solutions are in agreement with each other, from now on we use the parameters obtained from orbfit-lib.

\begin{table}
\centering
\caption{Parameters of the outer orbit.} 
\label{tab:long_orbit}            
\begin{tabular}{l l l}    
\hline\hline \noalign{\smallskip}           
Parameter & \multicolumn{2}{c}{Value}\\ 
 & orbfit-lib & spinOS \\
\hline \noalign{\smallskip}
$P$ (d) & $1603.24\pm0.19$ & $1603.54 \pm 0.25$ \\
$t_0$ (RJD) & $57300.02\pm0.57$ & $57299.9 \pm 1.2$ \\
$e$ & $0.5291\pm0.0011$ & $0.5261 \pm 0.0014$ \\
$a$ (mas) & $12.427\pm0.016$ & $12.368 \pm 0.067$\\
$i$ ($\degree$) & $78.216\pm0.030$ & $78.171 \pm 0.042$ \\
$\omega$ ($\degree$) & $300.43\pm0.08$ & $300.59 \pm 0.17$ \\ 
$\Omega$ ($\degree$) & $256.87\pm0.07$ & $256.834 \pm 0.056$ \\
$K_{1+2}$ (\SI{}{\km\per\s}) & $24.37\pm0.10$ & $24.26 \pm 0.35$ \\
$K_3$ (\SI{}{\km\per\s}) & $54.42\pm0.08$ & $54.24 \pm 0.18$ \\
$v_\gamma$ (\SI{}{\km\per\s}) & $3.03\pm0.04$ & $3.15 \pm 0.31$ \\
\noalign{\smallskip} \hline  \noalign{\smallskip}     
$M_\mathrm{tot}$ ($\Msun$) & $53.04 \pm 0.29$ & $52.72 \pm 0.81$ \\
$a$ (au) & $10.065 \pm 0.018$ & $10.054 \pm 0.051$ \\
$M_{1+2}$ ($\Msun$) & $36.63 \pm 0.18$ & $36.43 \pm 0.58$ \\
$M_3$ ($\Msun$) & $16.41 \pm 0.12$ & $16.29 \pm 0.35$ \\
$q$ & $0.4479 \pm 0.0019$ & $0.4473 \pm 0.0066$\\
$d$ (pc) & $809.9 \pm 1.8$ & $812.9 \pm 1.6$ \\ \noalign{\smallskip}
\hline
\end{tabular}
\tablefoot{$P$ is the sidereal orbital period, $t_0$ is the epoch of periastron passage, $e$ is the eccentricity, $\alpha$ is the angular size of the semi-major axis, $i$ is the inclination of the orbit, $\omega$ is the argument of periastron, $\Omega$ is the longitude of the ascending node, $K_{1+2}$ is the RV curve amplitude of the inner binary's centre of mass, $K_3$ is the RV curve amplitude of the tertiary, $v_\gamma$ is the systemic velocity, $M_\mathrm{tot}$ is the total mass of the whole system, $a$ is the semi-major axis, $M_{1+2}$ is the total mass of the inner eclipsing binary, $M_3$ is the mass of the tertiary, $q = M_3/M_{1+2}$ is the mass ratio, and $d$ is the distance to the system.}
\end{table}

\subsection{Modelling the inner eclipsing binary with PHOEBE}\label{inner-orbit}

The main model of the close eclipsing binary was constructed with the Python package PHOEBE (PHysics Of Eclipsing BinariEs, version 2.4.16 \citep{Prsa2016, Conroy2020}. PHOEBE allows both forward and inverse modelling of the observables of eclipsing binary systems while using advanced physics. It includes several algorithms for estimation, optimisation, and sampling of variables. It is, however, computationally demanding, especially for time-dependent systems (due to e.g. apsidal precession). For this reason, we only used PHOEBE for the TESS photometry and spectra mentioned in Section 2.1. The FRAM and Hipparcos light curves and older sets of RVs published in the literature were used only to obtain an estimate of the apsidal precession rate.

First, we used the Nelder-Mead optimiser to refine the values of the elements separately for each dataset - four light curves and three RV curves corresponding to the three different spectrographs. We assumed that during the time covered by each individual dataset, the argument of periastron does not change its value. This is a reasonable assumption for the photometry data covering only several days, while in the case of the RV curves, the value of $\omega$ actually changes by a few degrees. The situation is furthermore complicated by the fact that due to the low eccentricity of the system, the apsidal precession is not well visible in the data. Despite that, for each of the datasets, we were able to obtain a different value of $\omega$. Based on those, we estimated the rate of the apsidal advance, which we then used as an input value for KOREL, in order to obtain higher-quality RVs.

The Rossiter–McLaughlin effect was not included in the calculation, as it is not actually visible in the RVs obtained from KOREL. This allowed for a much faster computation. Since all components of $\delta$~Cir are hot stars, we set the irradiation coefficient as well as the gravity brightening (or darkening) coefficient to 1 for both the primary and the secondary. Because in PHOEBE there is no better option for hot stars, we used only blackbody atmospheres. Due to that, we had to manually set the values of the limb-darkening (LD) coefficients. For both stars, we used a linear LD coefficient of 0.35 and a logarithmic LD coefficient of 0.23. Rapid changes of unknown origin can be clearly seen in the light curve outside of the eclipses. We did not fit the limb darkening coefficients because the impact of rapid changes on the shape of the light curve is more important than the limb darkening.

Due to the light travel-time effect (LTTE), we had to fit the times of the primary minimum for each set of photometry separately. The LTTE in our system can cause a shift in the primary minimum of up about \SI{1.3}{\hour}, which is a non-negligible fraction of the 3.9~day orbital period. LTTE is treated properly in KOREL, so we do not have to account for it in our RV data. Since the shift in the time of the primary minima is caused only by the LTTE, we manually shifted the light curves in time, so that the times of the primary minima align with the times of superior conjunction for the RV curves. We fitted all of the RVs and both TESS light curves at once, and the model also included apsidal precession.

\begin{figure}
\centering
	\resizebox{\hsize}{!}{\includegraphics{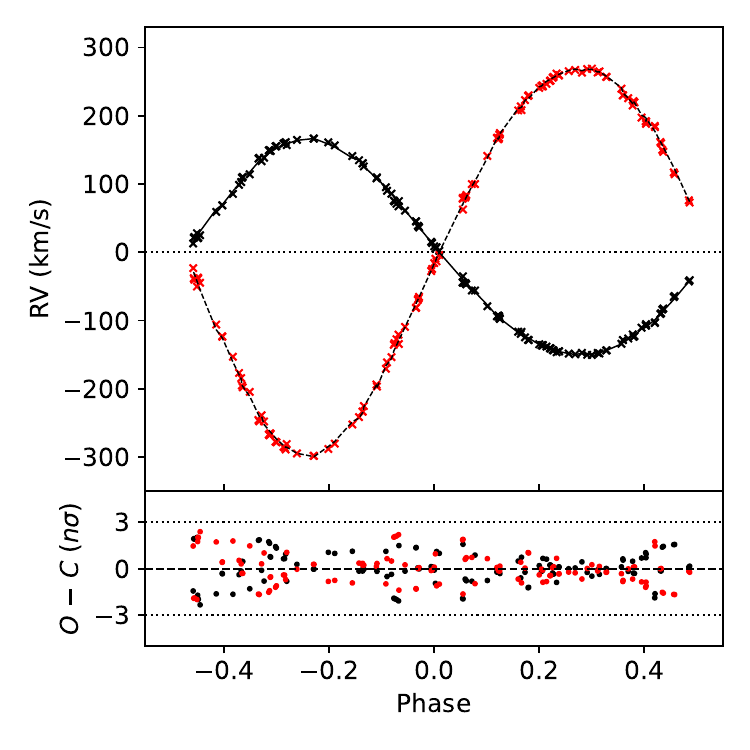}}
\caption{\textit{Upper:} RV curve of the inner orbit. Black x-signs denote the RVs of the primary, and the red x-signs denote the RVs of the secondary. The error bars are too small to be visible. The systemic velocity, $v_\gamma$, has been subtracted from the RVs. \textit{Lower:} $O-C$ residuals in the units of $\sigma$ for each individual point.}
\label{fig:mcmc_RV}
\end{figure}

% MB suggested
\paragraph{MCMC.}
After getting an initial model which agreed well with the data, we sampled the parameter space via MCMC.
We sampled 15 parameters, used 50 walkers, and did 1000 iterations. The sampled parameters include sidereal orbital period $P$, projection of semi-major axis $a \sin i$, eccentricity $e$, mass ratio $q$, argument of periastron $\omega$, rate of apsidal precession $\dot \omega$, inclination $i$, time of the primary minimum $t_{\mathrm{min}}$, ratio of effective temperatures $T_{\mathrm{eff,2}}/T_{\mathrm{eff,1}}$, radius of the primary $R_{1}$, radius of the secondary $R_{2}$, and, separately for each photometry dataset, also passband luminosity of the primary $L_{\mathrm{plum}}$ and the third light $l_3$. The absolute value $L_{\mathrm{plum}}$ is scaled in such a way that the luminosity of the whole system is $4\pi$ and the flux outside of the eclipses is roughly equal to 1. The third light is expressed as a fraction of the total luminosity made up by the third component. Since we verified the value of $v_\gamma$ in our fit of the outer orbit, we set it as a constant in our PHOEBE model and did not include this parameter in MCMC. The main reason to do this was to shorten the computation time.

\begin{figure*}
\centering
	\includegraphics[scale=0.7]{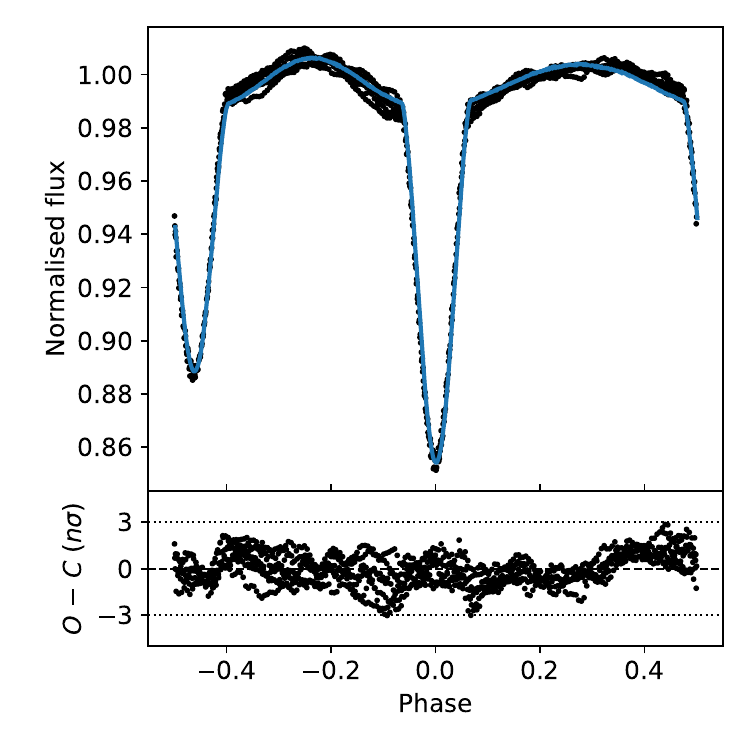}    
 	\includegraphics[scale=0.7]{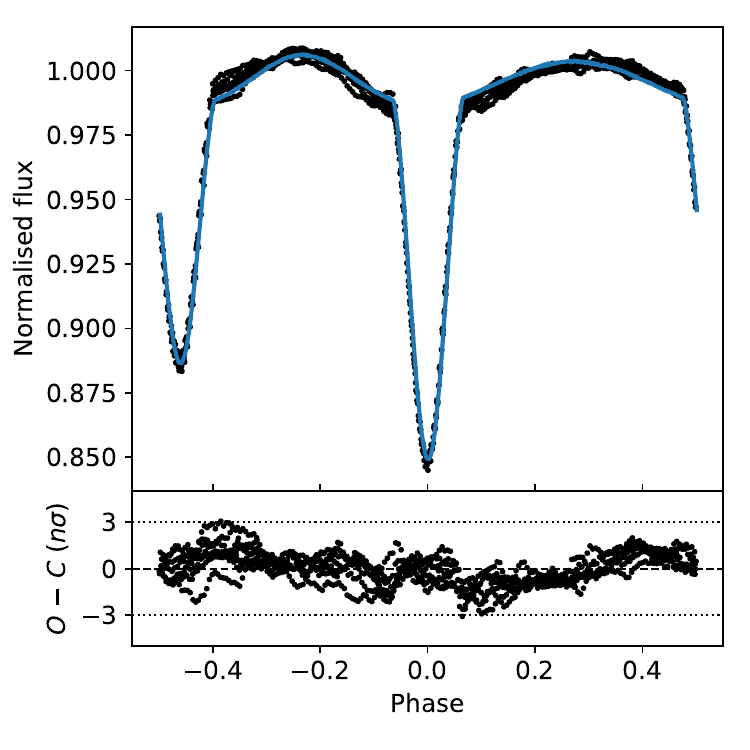}
\caption{Final phase plot of the TESS light curves from sector 12 and sector 65. \textit{Left:} light curve from TESS sector 12 and below its residuals in the units of $\sigma$. \textit{Right:} light curve and its residuals from TESS sector 65.}
\label{fig:mcmc_LC}
\end{figure*}

 The fitted parameters and their $1\sigma$ uncertainties are summarised in Table \ref{tab:short_orbit}. We drew 25 samples from the final MCMC distribution and calculated the models in Figs.~\ref{fig:mcmc_RV} and \ref{fig:mcmc_LC}.
The orbital solution obtained with PHOEBE provides a precise snapshot of the current configuration of $\delta$~Cir. Moreover, the physical properties derived from this model, such as component masses, radii, and rotation, also offer valuable constraints on the system’s evolutionary stage and future development.

\begin{table}[ht]
\caption[]{Orbital parameters of the short orbit, calculated for the epoch RJD 59625.66733.}
\label{tab:short_orbit}
\begin{center}
\begin{tabular}{lr}
\noalign{\smallskip}\hline\hline\noalign{\smallskip}
Parameter & Value  \\
\noalign{\smallskip}\hline\noalign{\smallskip}
$P$ (d) & $3.90244719 \pm 0.00000052$\\ 
$a \sin i$ ($\Rsun$) & $33.895 \pm 0.011$\\ 
$e$ & $0.06116 \pm 0.00010$\\ 
$q$ & $0.55655 \pm 0.00064$\\ 
$\omega$ (\degree) & $344.83 \pm 0.37$\\ 
$\dot \omega$ (\degree/ yr) & $1.822 \pm 0.027$\\ 
$i$ (\degree) & $78.7745 \pm 0.0088$\\ 
$t_{\mathrm{min}}$ (RJD) & $58650.055430 \pm 0.00010$\\
$T_{\mathrm{eff,2}}/T_{\mathrm{eff,1}}$ & $0.7723 \pm 0.0013$\\ 
$R_{1}$ ($\Rsun$) & $9.647 \pm 0.013$\\ 
$R_{2}$ ($\Rsun$) & $5.3997 \pm 0.0076$\\ 
$L_{\mathrm{plum,S12}}$ (a.u.) & $7.279 \pm 0.017$\\ 
$L_{\mathrm{plum,S65}}$ (a.u.) & $7.431 \pm 0.011$\\ 
$l_{\mathrm{3,S12}}$ & $0.2990 \pm 0.0010$\\ 
$l_{\mathrm{3,S65}}$ & $0.28419 \pm 0.00074$\\ \noalign{\smallskip}
\hline  \noalign{\smallskip}     
$M_{1+2}$ ($\Msun$) & $36.355 \pm0.036 $ \\
$M_1$ ($\Msun$) & $23.358 \pm 0.030$ \\
$M_2$ ($\Msun$) & $12.997 \pm 0.019$ \\
$\log g_1$ & $3.8375\pm0.0013$\\
$\log g_2$ & $4.0871\pm0.0015$\\
\noalign{\smallskip}\hline
\end{tabular}
\end{center}
\end{table}

\section{Evolutionary models with MESA}\label{MESA}

To explore the future evolution of $\delta$~Cir, we employed the 1D stellar evolution code MESA \citep[Modules for Experiments in Stellar Astrophysics;][]{Paxton2011ApJS..192....3P,Paxton2015ApJS..220...15P,Paxton2018ApJS..234...34P,Paxton2019ApJS..243...10P}, using its modules for both single-star and binary evolution.
MESA evolves stellar models in discrete time steps, solving at each step the coupled stellar structure and composition equations via a Newton–Raphson scheme. The solver iteratively refines the solution until the desired level of convergence is achieved.

We evolved $\delta$~Cir from the pre-main sequence all the way to core collapse. We assumed the solar metallicity $Z = 0.014$ \citep{Asplund2009ARA&A..47..481A} and considered rotation and stellar winds. 

\subsection{Single-star tracks}
\begin{figure}
\centering
\includegraphics[width=0.49\textwidth]{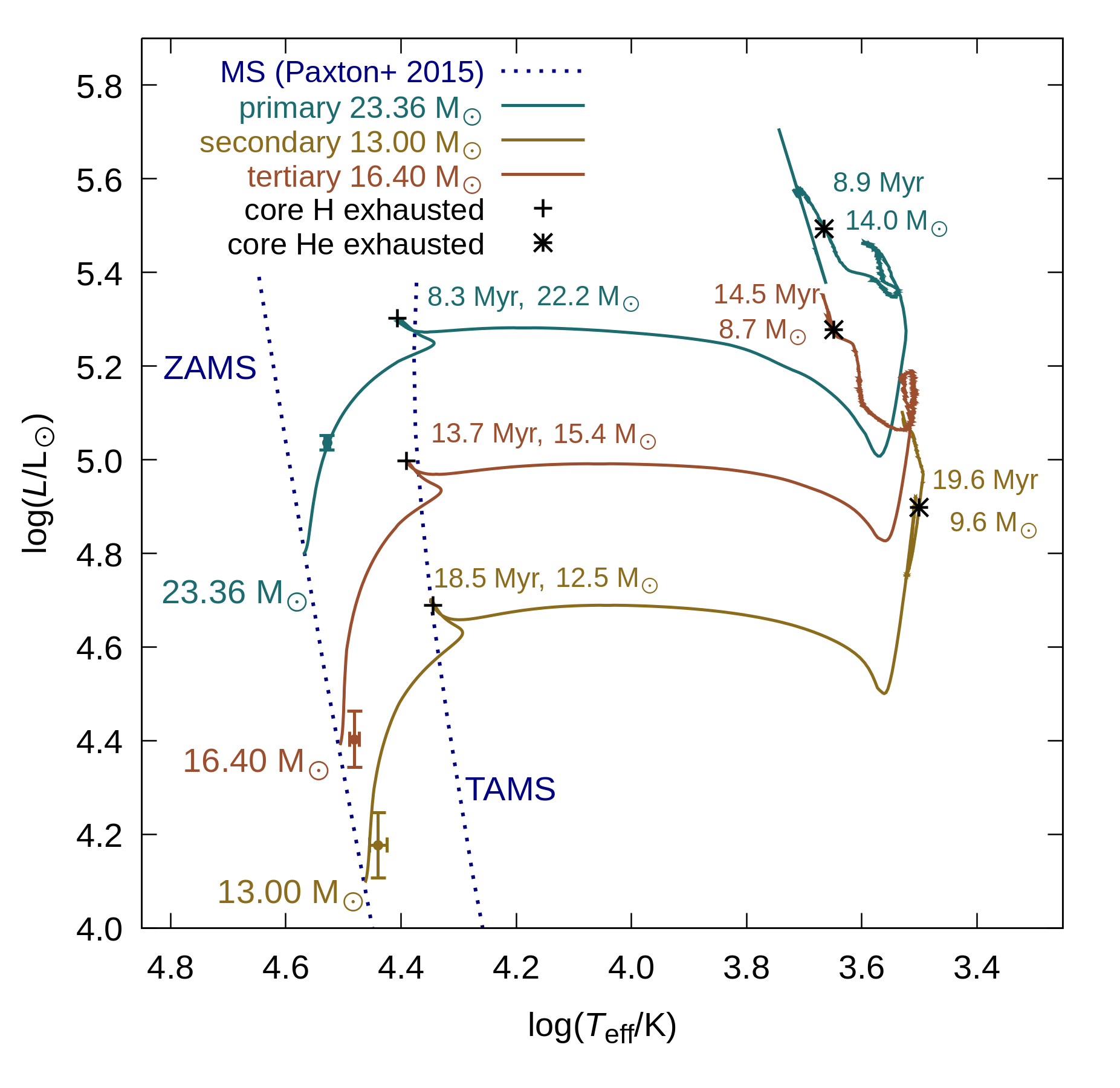}
\caption{ HR diagram with evolutionary tracks computed with MESA for single-star models with masses corresponding to the components of $\delta$~Cir. The tracks begin at the zero-age main sequence (ZAMS) and end at core collapse. 
The dotted lines denote the ZAMS and the terminal-age main sequence (TAMS) for single, non-rotating stars without stellar wind mass loss \citep{Paxton2015ApJS..220...15P}.
Stellar winds and rotation are included in our models. The observed positions of all three stars are shown as dots with error bars. The primary has evolved away from the ZAMS, while the secondary and the tertiary remain close to it. Markers indicate when hydrogen and helium are exhausted, and the adjacent labels show the corresponding ages and masses.
}
\label{HRD_delCir_single}
\end{figure}

First, we compare the resulting HRD positions of all components with single-star evolutionary tracks (see Fig.~\ref{HRD_delCir_single}). From the single-star evolutionary model of the tertiary, we estimate its age to be $(3.8 \pm 1.3)$\,Myr based on the derived uncertainty of rotational velocity and mass, and assuming a 10\% uncertainty in radius. According to the model, the tertiary will evolve into a supergiant before core collapse, moving towards cooler temperatures and slightly higher luminosities. Core helium exhaustion is expected to occur in about 10\,Myr, by which time the star will have lost almost half of its present mass.
If both the primary and the secondary were to evolve as isolated stars, they would also move towards cooler effective temperatures and higher luminosities, with the primary becoming a supergiant and the secondary a giant. In reality, however, the close proximity of the two components means that their subsequent evolution will be strongly coupled, likely involving mass transfer and interaction effects that will significantly alter their single-star evolutionary paths.

\subsection{Binary tracks}
\begin{figure}
\centering
\includegraphics[width=0.49\textwidth]{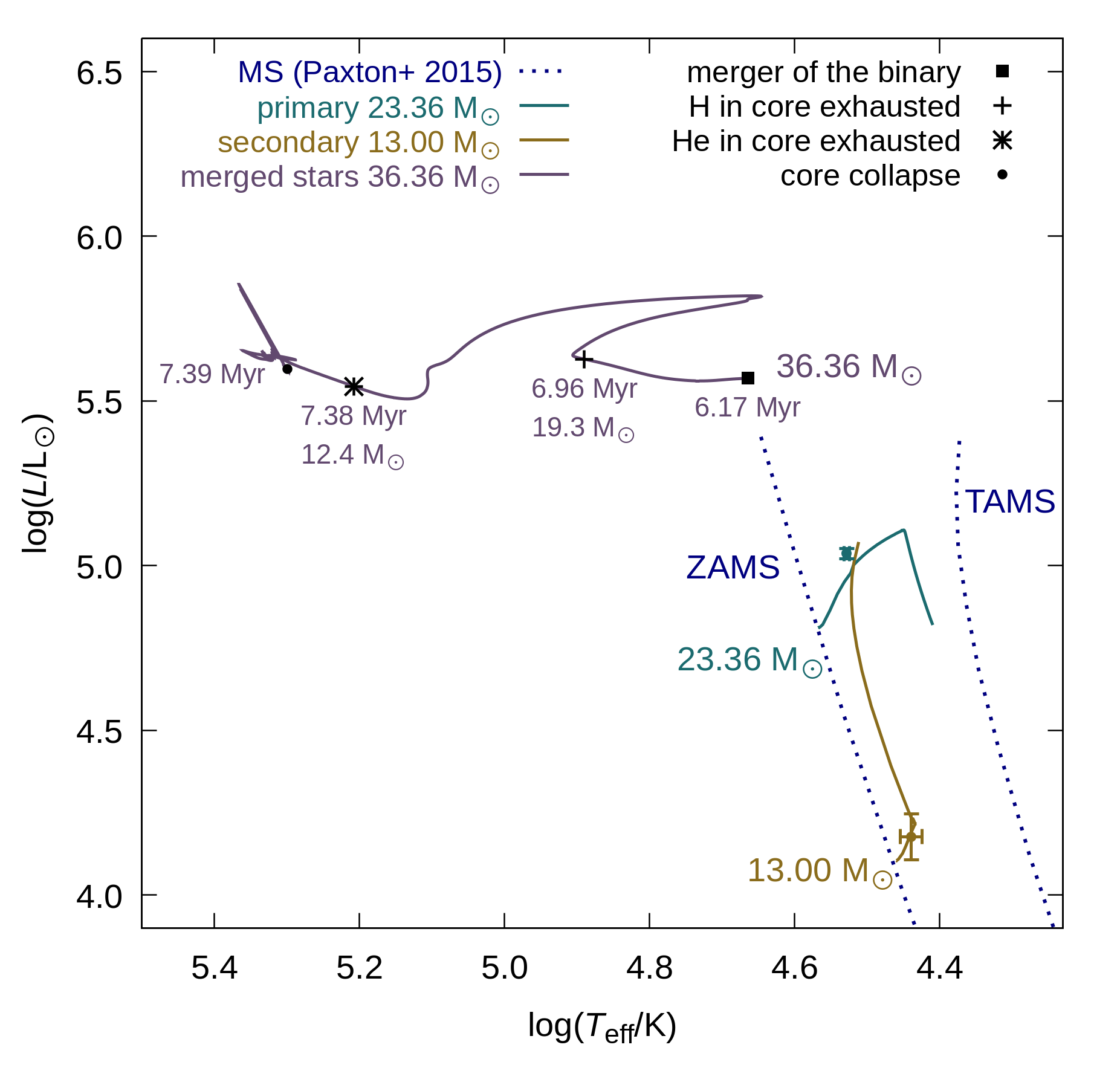}
\caption{ Same as Fig.~\ref{HRD_delCir_single} but with evolutionary tracks for the inner binary (primary and secondary) computed from the ZAMS until the onset of Roche lobe overflow by the accretor. Subsequently, we assumed a common-envelope phase and a merger resulting in a star with a mass of 36.36\,\Msun \, rotating at 85\% of its critical velocity (typical velocity for mergers) and having a stellar wind with a Dutch scaling factor of 1.0. 
This star evolves into a hotter WR star, unlike the more slowly rotating and less massive stars in Fig.~\ref{HRD_delCir_single}, which evolve into supergiants or giants. Based on the compactness parameter, the core collapse is expected to result in a black hole.
}
\label{HRD_delCir_binary}
\end{figure}

For the evolution of the close binary (primary and secondary), 
we evolved both stars using the binary module and assuming a conservative mass transfer scheme \citep{Kolb1990A&A...236..385K}.
Again, we included stellar winds and rotation set to the same initial values as for single-star evolutions.
Fig.~\ref{HRD_delCir_binary} shows the HRD with evolutionary tracks and the current positions of primary and secondary. According to the radii from our modelling, the components are coeval, and their current ages are $(4.4\pm 0.1)$\,Myr and $(4.7 \pm 0.2)$\,Myr for primary and secondary, respectively.

\begin{figure}
\centering
\includegraphics[width=0.49\textwidth]{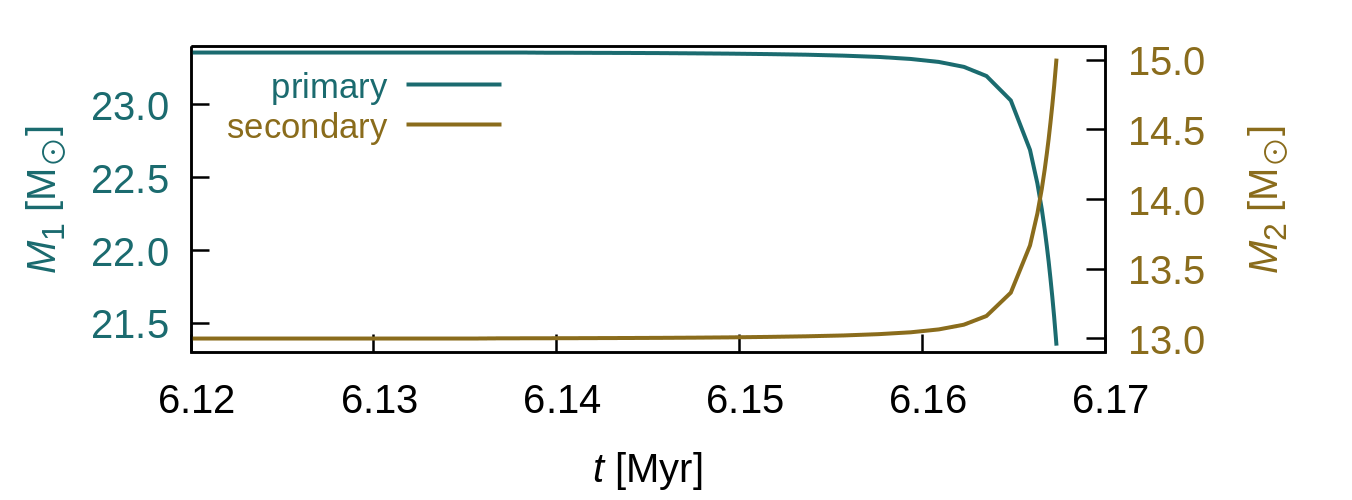}
\includegraphics[width=0.49\textwidth]{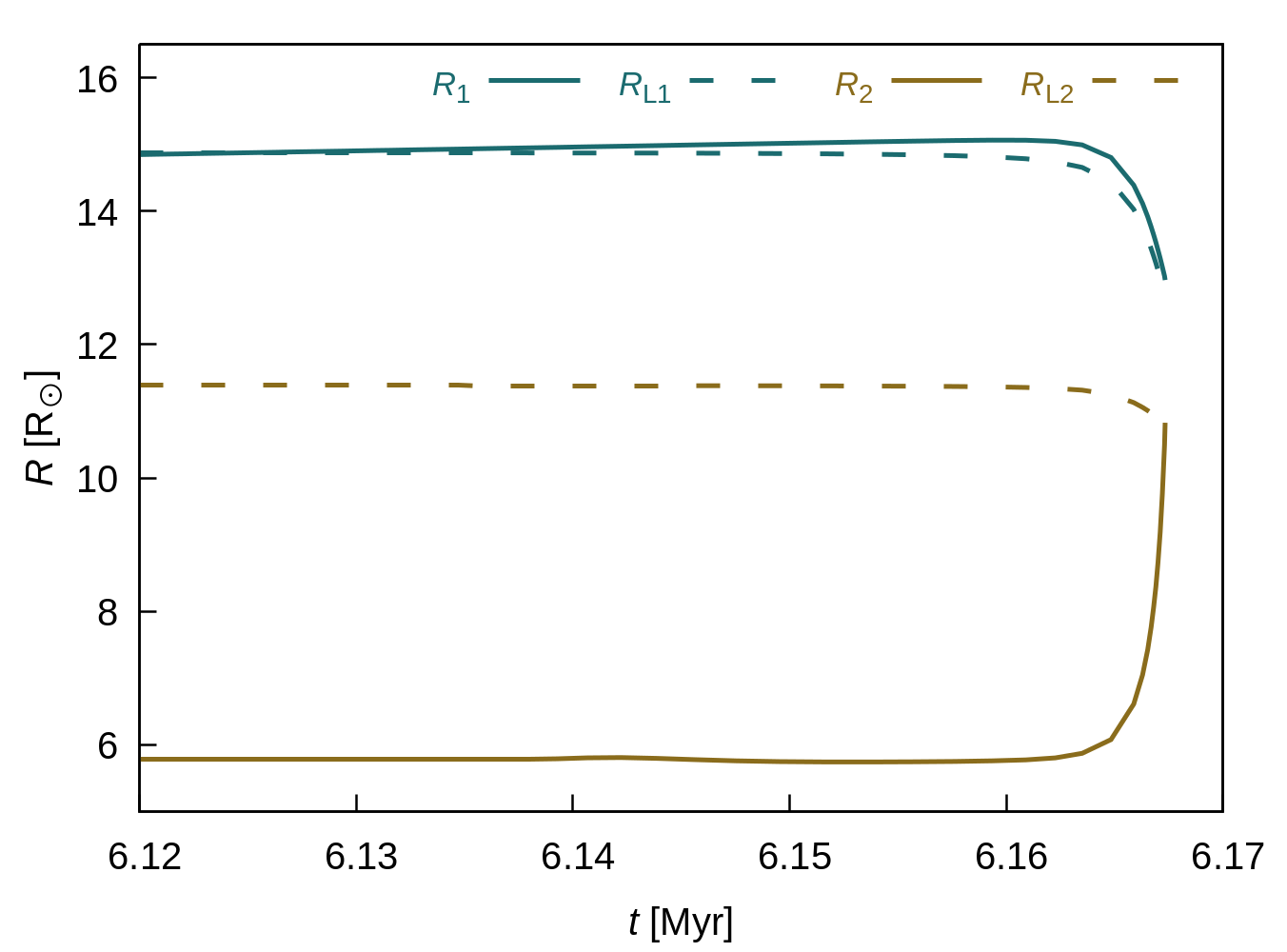}
\includegraphics[width=0.49\textwidth]{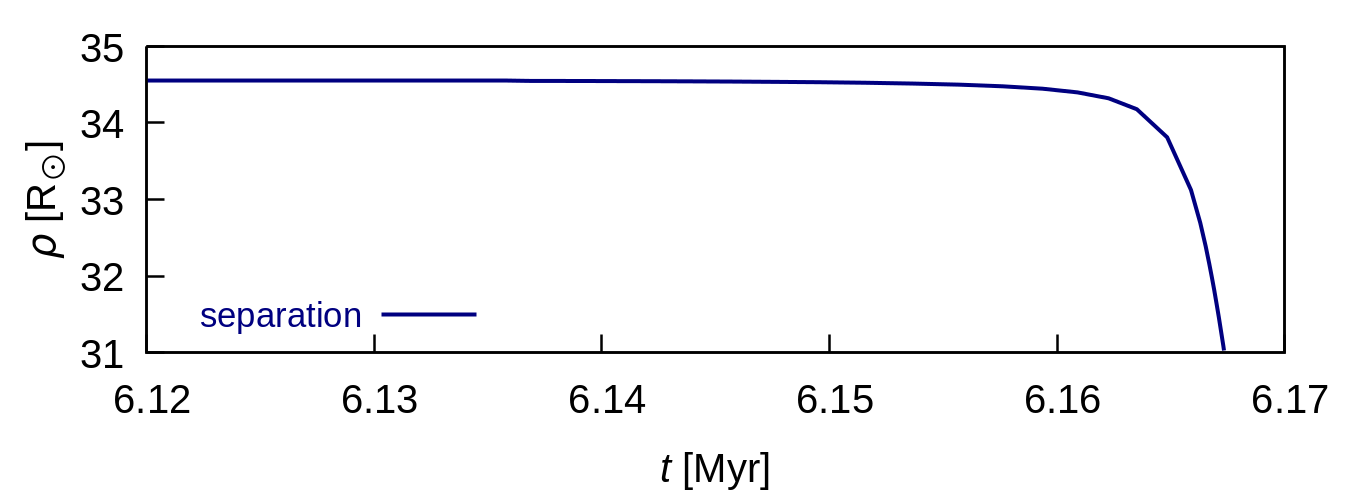}
\caption{
Evolution of the binary masses, radii, and orbital separation between the onset of mass transfer and the common-envelope phase. The mass transfer is dynamically unstable (runaway) because the donor overfills its Roche lobe more and more. Such mass transfer will lead to a common envelope phase and is likely to result in a merger.
}
\label{binary_mass_sep}
\end{figure}

\begin{figure}
\centering
\includegraphics[width=0.49\textwidth]{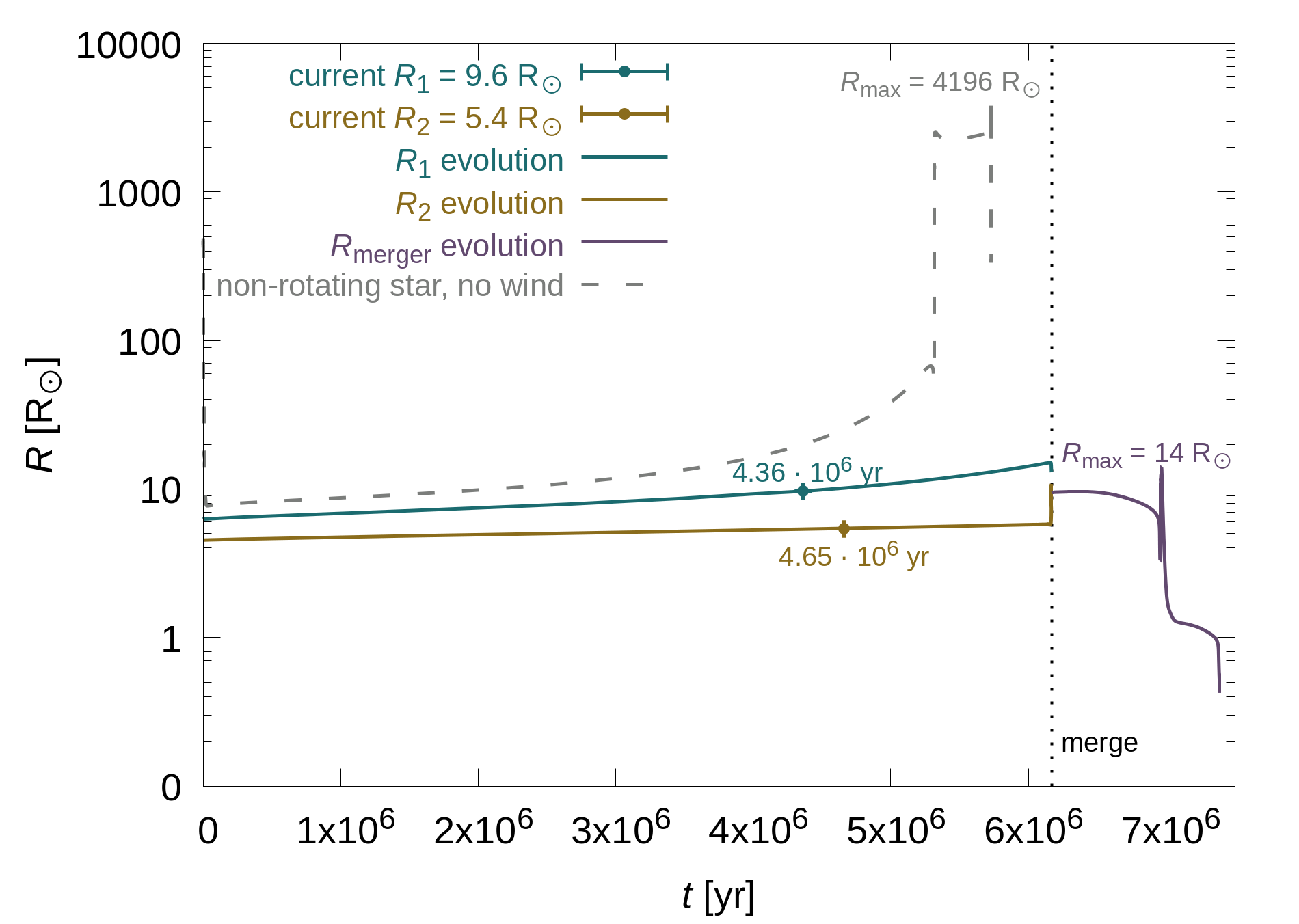}
\caption{ Evolution of the components’ radii in the eclipsing binary. The radii of the primary and the secondary stars are shown to evolve up to the phase of the common envelope. Afterwards, we illustrate the subsequent evolution of the merger product’s radius. Its maximum value is only a few solar radii, significantly smaller than that of a single, non-rotating star without stellar winds (dotted grey line). Such a small radius before the core collapse confirms the merger's position in the HR diagram, implying that it will evolve into a WR star and exert negligible influence on the tertiary companion.
}
\label{HRD_delCir_binary_radii}
\end{figure}

The evolution of the binary components is followed up to the moment when the gainer (secondary) overflows its Roche lobe, i.e. the overflow indicator, $\frac{(R - R_{\mathrm{L}})}{R_\mathrm{L}}$, is greater than zero.
This happens after the donor's radius, $R_1$, intersects and exceeds its Roche-lobe radius, $R_\mathrm{L1}$, and the mass transfer rate rapidly increases in the last 5000\,yr, see Fig.~\ref{binary_mass_sep}. 
The secondary could not thermally adjust to the accreted material and doubled its radius in a few thousand years to a value of 10.8\,\Rsun \ (see Fig.~\ref{HRD_delCir_binary_radii}).

The mass transfer is unstable because the system exhibits runaway behaviour. $R_2$ increases sharply, while $R_\mathrm{L2}$ does not respond accordingly, and $R_1$ and $R_\mathrm{L1}$ decrease. If the transfer were stable, the donor’s radius and the Roche-lobe radius would maintain a roughly constant ratio after the contact, resulting in smooth, self-regulating evolution. The orbital separation and period decrease sharply, by almost 5\,\Rsun \ and 0.6\,d over only 5000\,yr. This unstable mass transfer ultimately triggers a common-envelope phase.

\subsection{Common-envelope phase}
During the common-envelope phase, it is uncertain whether the contact binary will actually merge \citep{Pols1994A&A...290..119P,Podsiadlowski2010MNRAS.406..840P,Justham2014ApJ...796..121J}. Although both stars are expected to keep expanding as they evolve, and since they are already in contact, a merger is very likely.

MESA cannot simulate the merging process itself, nevertheless we continued the model by evolving the resulting merger product. We assumed that it has a total mass of $36.36\,\Msun$, a radius of $R\sim M^{0.62} \approx 9.5$\,\Rsun, and a high rotation rate (85\%) typical for mergers when the orbital angular momentum from the pre-merger system is largely converted into the spin of the merger product \citep{deMink2011IAUS..272..531D}.
From Figs.~\ref{HRD_delCir_binary} and~\ref{HRD_delCir_binary_radii}, we see that stellar winds and rapid rotation prevent the merger's envelope expansion typically observed in massive stars before core collapse. A potential merger did not enter a red supergiant phase (that would correspond to radii of thousands of \Rsun\ and the upper-right region of the HRD) and will instead evolve into a Wolf–Rayet star. Wolf–Rayet stars maintain small radii (a few solar radii) even near core collapse, never move to the cool side of the HRD, and occupy the upper-left region of the HRD, remaining hot and luminous before the core collapse. The merging of the close binary, therefore, will not reach the third component in the distance of $\approx 2172$\,\Rsun.

\subsection{Core collapse}
\citet{OConnor2011ApJ...730...70O} found that the outcome of core collapse can be estimated, to the first order, by a single parameter $\xi_{M}$, the compactness of the stellar core at bounce

\begin{equation}
    \xi_{M} = \frac{M/\Msun}{R(M_\mathrm{bary} = M)/1000\,\mathrm{km}}\Bigg\rvert_{t=t_\mathrm{bounce}},
\end{equation}
where $M$ is the chosen mass coordinate inside the star and $R$ is the radius that surrounds that mass. Neutrino-driven supernova explosions can be launched up to a bounce compactness $\xi_{2.5} \lesssim 0.45$ (less compact cores), while the formation of black holes without explosion as outcomes is for very compact cores with $\xi_{2.5} \gtrsim 0.45$. 

We calculated the compactness parameter $\xi_{2.5}$ from the MESA pre-collapse profile (the last one).
The last profile for the potential merger product of the close binary shows that the core of the mass 2.5\,\Msun \ has a radius of
4407\,km. It results in the compactness parameter 0.567, indicating that such a merger will become a black hole.

\section{Discussion}

Since we obtained a lot of data covering both the outer and the inner orbit, our models of this system are overdetermined. Therefore, we can do several cross-checks of parameters obtained from different methods. The total mass of the eclipsing binary determined in PHOEBE is in good agreement with the one obtained from fitting the outer orbit. If we assume synchronous rotation of the primary and secondary, we can calculate the radii from the projection of their rotational velocity $v_\mathrm{r} \sin i$ in our PYTERPOL fit of the disentangled lines. For the primary, we get $R_1 = (9.2 \pm 0.4)\, \Rsun$, which is in good agreement with the radius from PHOEBE. However, for the secondary, we get $R_2 = (7.5 \pm 0.8)\, \Rsun$, while the one from PHOEBE is $5.4\, \Rsun$. We consider the values of radii from PHOEBE to be more reliable, so this likely means that the secondary does not rotate synchronously.

We can also compare the surface gravity and temperatures obtained from fitting the disentangled line profiles in PYTERPOL and from our model in PHOEBE. As seen in the Tables \ref{tab:pyterpol} and \ref{tab:short_orbit}, the values of $\log g$ agree with each other for both the primary and the secondary. The temperature ratio determined in PYTERPOL ($0.815\pm0.031$) is slightly higher than the value obtained in PHOEBE ($0.7723\pm0.0013$).

The theoretical value of the apsidal motion period driven by the third body is around 4700 yr, while the observed timescale is only 200\,yr. We can thus conclude that the apsidal motion is caused mostly by the tidal effects in the eclipsing binary and the impact of the third body is quite small.

The system is dynamically very stable; it is far from the stability criterion set by \citet{Mardling2001}. The two orbits are coplanar, so the Kozai-Lidov effect does not play a role in its dynamical evolution. The temperatures, masses and radii of all three components are in good agreement with the stellar parameters of main sequence stars determined from detached binaries compiled by \citet{Harmanec1988}. The luminosities of the primary and secondary can easily be calculated from their radii and effective temperatures. The luminosity of the third star can be calculated in several independent ways: from the flux ratios in interferometry, from the third light in the TESS and FRAM light curves, and also from the relative luminosity obtained when fitting the spectral lines in PYTERPOL. All the methods combined except for the TESS light curves give us the third light $l_3 = (0.175\pm0.022)$, but the value determined from the TESS light curves is significantly higher - it almost reaches 0.3.  Since $\delta$ Cir is very bright, it saturates the CCD pixels in the centre of the PSF, where each pixel measures \ang{;;21} on the side. $\delta$ Cir spans up to 7 $\times$ 5 pixels on the CCD in Sector 12 and 8 $\times$ 5 pixels in Sector 65, which corresponds to \ang{;2.45;} $\times$ \ang{;1.75;} (S12) and \ang{;2.8;} $\times$ \ang{;1.75;} (S65) on the sky. $\delta$ Cir is located only $3\degree$ from the galactic plane, in a field densely populated by stars, which can be seen for example in images from the VVV eXtended ESO Public Survey at the VISTA telescope in the $H$, $J$, and $K_s$ bands. According to the ASAS3 Catalogue in the $V$ band \citep{Pojmanski2004}, there are several objects within close proximity of Del Cir, whose $V$ band magnitudes range from about 8 to 12. The large difference in the third light could be caused by the contribution of light from these nearby sources.

We also tried to reproduce the observed stellar properties with the Bayesian stellar evolution tool BONNSAI\footnote{The BONNSAI web-service is available at \url{www.astro.uni-bonn.de/stars/bonnsai}.}\citep{Schneider2014}. We provided the observed values of $T_\mathrm{eff}$, $\log L$, and $v_\mathrm{r} \sin i$, assumed flat priors for the initial rotational velocity and age, and used the Salpeter initial mass function. The results are summarised in Table \ref{tab:bonnsai}. We were able to successfully reproduce the observed stellar parameters of both the primary and the secondary. However, the calculated mass and especially $\log g$ of the third star are different from the observed values. The value of $\log g$ obtained in PYTERPOL could be influenced by the renormalisation of the disentangled spectrum. The ages derived by BONNSAI suggest that all three stars are coeval and are also in agreement with the ages determined from our MESA model; we refer to Sect.~\ref{MESA}.

\begin{table}
\centering  
\caption{Stellar parameters obtained from the BONNSAI tool.} 
\label{tab:bonnsai}              
\begin{tabular}{l l l l}    
\hline\hline \noalign{\smallskip}          

Parameter & Primary & Secondary & Tertiary\\
\hline \noalign{\smallskip}
$M_{\mathrm{ini}}$ ($\Msun$) & $24.20^{+0.29}_{-0.45}$ & $12.60^{+0.62}_{-0.66}$ & $15.40^{+0.45}_{-0.54}$ \\  \noalign{\smallskip}
Age (Myr) & $4.22^{+0.18}_{-0.19}$ & $6.26^{+2.41}_{-2.96}$ & $4.3^{+1.1}_{-1.4}$ \\  \noalign{\smallskip}
$M$ ($\Msun$) & $23.60^{+0.36}_{-0.31}$ & $12.60^{+0.62}_{-0.66}$ & $15.40^{+0.45}_{-0.54}$ \\  \noalign{\smallskip}
$R$ ($\Rsun$) & $9.62^{+0.28}_{-0.24}$ & $5.26^{+0.54}_{-0.51}$ & $5.79^{+0.46}_{-0.42}$ \\  \noalign{\smallskip}
$\log g$ & $3.84^{+0.03}_{-0.02}$ & $4.10^{+0.08}_{-0.10}$ & $4.09^{+0.06}_{-0.06}$ \\  \noalign{\smallskip}
\hline                    
\end{tabular}
\end{table}

\begin{figure}
  \resizebox{\hsize}{!}{\includegraphics{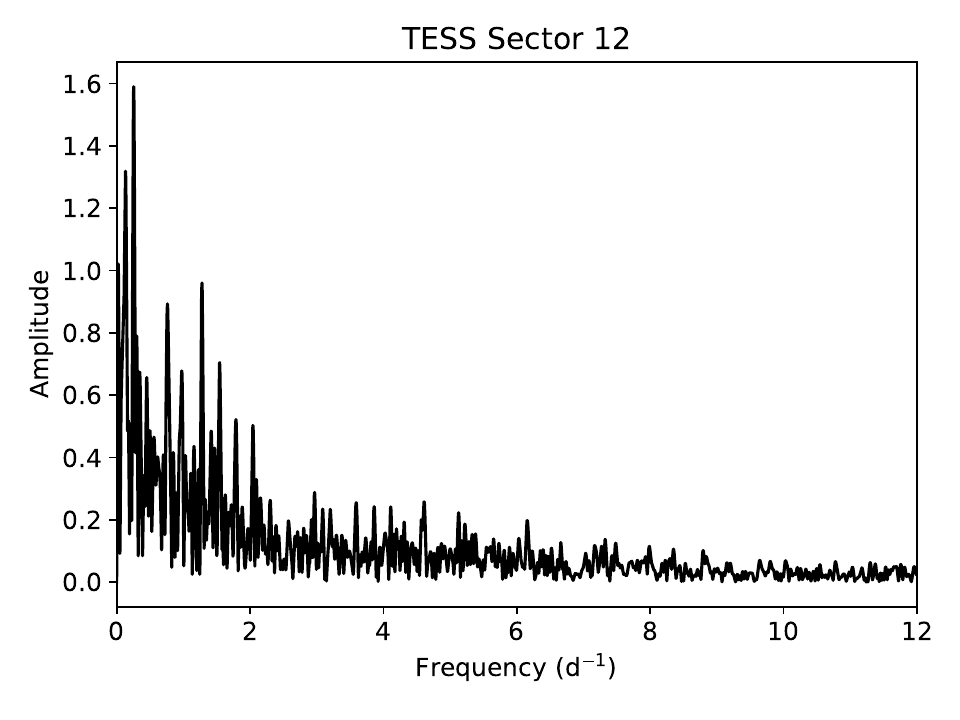}}
  \resizebox{\hsize}{!}{\includegraphics{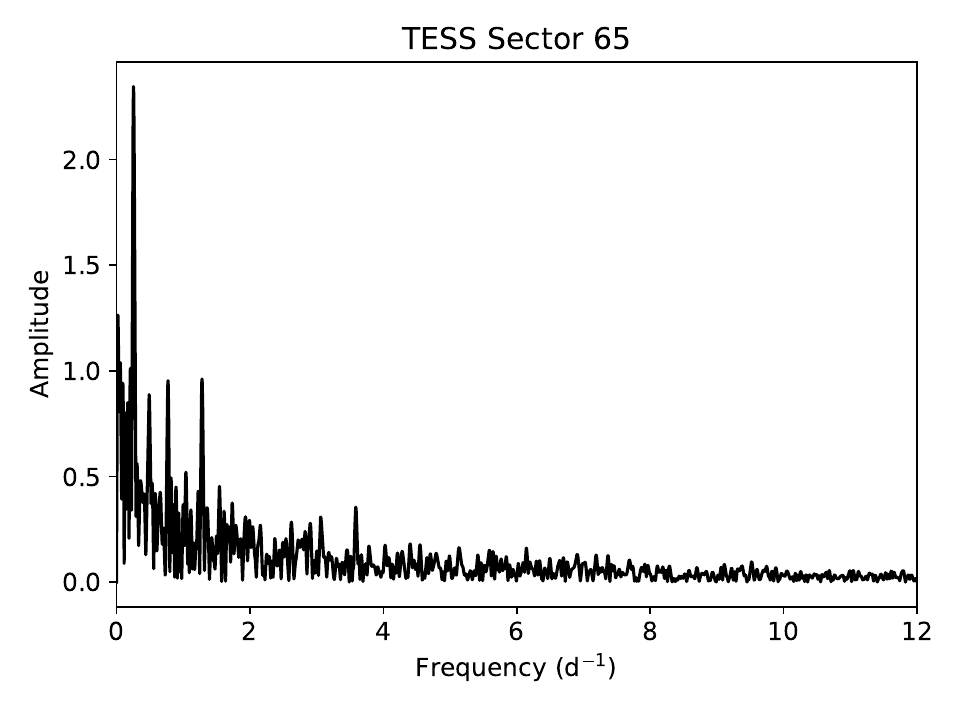}}
  \caption{Frequencies in light curve residuals. The only significant frequency peak in the upper plot at $f\sim0.254\, \mathrm{d}^{-1}$ corresponds to the 3.9~d orbital period. This is caused by the imperfect fit of the light curves.}
  \label{fig:frequencies}
\end{figure}

To see whether the rapid changes visible in the TESS light curves could be caused by pulsations, we searched for possible frequencies present in the light curve residuals. We calculated the periodograms for both TESS sectors using the Period04 code \citep{Lenz2004}. The result is shown in Fig.~\ref{fig:frequencies}. The only significant frequency peak (with signal-to-noise ratio > 5) was found in the Sector 12 residuals at $f\sim0.254\, \mathrm{d}^{-1}$. This frequency corresponds to the 3.9~d orbital period, which is an effect of imperfect fit. No other significant frequencies were found. Both periodograms show stochastic low-frequency variability, which is commonly observed in light curves of early-type stars. This has already been reported for the $\delta$ Cir light curve from TESS Sector 12 by \citet{southworth2022}. 

The distance to the system derived from the \textit{Gaia} DR3 parallax is $860^{+410}_{-210}$ pc, where we applied zero point and external uncertainty parallax corrections to the catalogue values\footnote{\url{https://vizier.cds.unistra.fr/viz-bin/VizieR?-source=I/355&-to=2}} according to \citet{2022A&A...657A.130M} and used a prior for OB-type stellar distribution following \citet{Pantaleoni2025} for the Bayesian inference of the distance from the parallax. For the final value and its uncertainties given above, we used the median of the posterior distribution and the 1$\sigma$ equivalent percentiles. As expected, the uncertainties are very large due to $\delta$~Cir being a very bright star and a multiple low-contrast system. The distance based on the orbital fit of the long orbit is superior in terms of precision by more than two orders of magnitude at (809.9 $\pm$ 1.8) pc. This distance would put it close to the centre of a subgroup of the Circinus complex called ASCC 79. With the total mass over 50 \Msun, $\delta$ Cir alone can play an important role in ASCC 79, since according to \citet{Kerr2025} it can make up to 10\% of the total mass of the group.

\section{Conclusions}

Determining all basic physical properties of massive OB-type stars is important for constraining theoretical evolutionary models. The stellar parameters can be directly determined from the orbital solutions of binary stars. Among the massive OB stars, the hierarchical triple star $\delta$~Cir represents an ideal laboratory for this purpose. We combined spectroscopic, photometric, and interferometric observations to obtain a complete model of this system (see Tables~\ref{tab:long_orbit}, \ref{tab:short_orbit}). 
Using the VLTI/PIONIER and GRAVITY observations, we constrained the astrometric orbit of the third body and the eclipsing binary for the first time. This allowed us to disentangle several spectral lines, which we then used to obtain the radiative properties of all three components (see Table \ref{tab:pyterpol}). The primary and tertiary are late O-type stars, while the secondary is an early B-type star. Both orbits lie in the same plane, and their period ratio is quite large; the system is dynamically very stable in the long term, and no complicated three-body dynamical effects occur in the system. The slightly eccentric inner orbit undergoes apsidal rotation with a period of about 200 years. No mass transfer has occurred in the system so far: all three stars are evolving as single stars, and we found their physical parameters in good agreement with the theoretical evolutionary models. Based on the MESA evolutionary models, a dynamically unstable mass transfer in the inner binary is predicted to occur at an age of 6.1 Myr. The subsequent common-envelope phase will lead to the merger of the binary, forming a single object with a mass of approximately 36.4\,\Msun. This rapidly rotating, mass-losing star will evolve into a Wolf–Rayet star and eventually collapse into a black hole. The current ages of the components are estimated to lie between 3.8 and 4.7\,Myr.
The precise distance to the system $(809.9 \pm 1.8)$ pc, derived from the angular and absolute size of the outer orbit, suggests that $\delta$~Cir belongs to a young stellar population ASCC 79, and it is by far the most massive member of this group.

\begin{acknowledgements} Based on observations collected at the European Southern Observatory under ESO programmes 089.C-0211(A), 189.C-0644(A), 093.C-0503(A), 098.D-0706(B), 099.D-0777(B), 596.D-0495(D), 596.D-0495(J), 5100.D-0721(D), 109.23HT.001, 182.D-0356(B), 182.D-0356(E), 185.D-0056(A), 185.D-0056(B), 185.D-0056(C), 185.D-0056(E), 185.D-0056(I), 185.D-0056(K), 	178.D-0361(B), 178.D-0361(D), 178.D-0361(F), 198.B-2004, on photometry collected by the TESS mission and obtained from the MAST data archive at the Space Telescope Science Institute (STScI), on photometries from the Hipparcos and FRAM instruments, and on spectra from the CTIO CHIRON echelle spectrograph.
This research has made use of the Jean-Marie Mariotti Center Optical Interferometry Database\footnote{OIDB available at \url{http://oidb.jmmc.fr}.}. We acknowledge the use of programmes KOREL, written by Petr Hadrava, and reSPEFO, written by Adam Harmanec.
We would also like to thank the Pierre Auger Collaboration for the use of its facilities. The operation of the robotic telescope FRAM has been supported by the grants of the Ministry of Education of the Czech Republic (MSMT-CR LG13007 and LG15014) and by the grant No. 14-17501S of the Czech Science Foundation.
This research has made use of NASA's Astrophysics Data System Bibliographic Services and the SIMBAD database, operated at CDS, Strasbourg, France. A.O. was supported by GA UK grant no. 113224 of the
Grant Agency of Charles University.
\end{acknowledgements}

\bibliographystyle{aa}
\bibliography{delcir}

\begin{appendix}
\section{Notes on the MESA model}
For our MESA models, we used the \texttt{approx21-cr60-plus-co56.net} nuclear network, which includes isotopes up to iron and cobalt. 
We adopted the iron-core infall limit as the stopping condition, which defines the threshold inward velocity of the stellar core that MESA uses to decide when a star has reached core collapse.

We ran a continuous sequence of phases: pre-MS contraction,
H-burning, He-burning, advanced burning (C, Ne, O, Si), and
core collapse.  
We set the rotation rates of the primary, secondary, and tertiary to 17\%, 14\%, and 21\% 
of the critical rotation speed at the ZAMS, respectively. The values were estimated by fitting the spectral lines in PYTERPOL.
To involve stellar winds into our models, we used the 'Dutch' scheme for hot and cold winds \citep{Glebbeek2009A&A...497..255G}. 
Depending on the star’s effective temperature and surface
hydrogen fraction, the 'Dutch' scheme adopts prescriptions from \citet{Nugis2000A&A...360..227N, Vink2001A&A...369..574V} for hot winds and \citet{deJager1988A&AS...72..259D,Nieuwenhuijzen1990A&A...231..134N,vanLoon2005A&A...438..273V} for cold winds.
The typical Dutch scaling parameter for massive stars ranges between 0.5 and 1.5; we adopted a value of 1.0 for all components.

The compactness parameter $\xi_{M}$ that we used to estimate the fate of the merger is highly non-monotonic with the total mass of a~star \citep{OConnor2011ApJ...730...70O}.
This parameter depends only on the core structure, on the radius enclosing the given mass, which is influenced by mass loss (winds), convective history and burning stages, rotation, metallicity, and overshooting. \citet{Pejcha2015ApJ...801...90P} quantitatively tested how well compactness predicts supernova success or failure and arrived at the correctness of prediction of 88\% for the cores with masses between 1.7 and 2.2\,\Msun. 
\end{appendix}

\end{document}